\documentclass[12pt,preprint]{aastex}
\usepackage{color}

\shorttitle{NGC~5253 H53$\alpha$ RRLs}
\shortauthors{Rodriguez-Rico et al.}

\begin{document}

\title{VLA H53$\alpha$ observations of the central region of the Super Star Cluster Galaxy NGC~5253}

\author{C. A. Rodr\'{\i}guez-Rico \altaffilmark{1,2}}
\email{carlos@astro.ugto.mx}

\author{W. M. Goss\altaffilmark{3}} \email{mgoss@nrao.edu}

\author{J. L. Turner\altaffilmark{4}} \email{turner@astro.ucla.edu}

\author{Y. G\'omez\altaffilmark{2}} \email{y.gomez@astrosmo.unam.mx}

\altaffiltext{1}{Departamento de Astronom\'{\i}a, Universidad de Guanajuato, Guanajuato,
Guanajuato, M\'exico}

\altaffiltext{2}{Centro de Radioastronom\'{\i}a y Astrof\'{\i}sica,
UNAM, Campus Morelia, Apdo. Postal 3-72, Morelia, Michoac\'an 58089,
M\'exico.}

\altaffiltext{3}{National Radio Astronomy Observatory, Socorro, NM, U.S.A.  87801}

\altaffiltext{4}{Department of Physics and Astronomy, UCLA, Los Angeles, CA 90095 USA.}

\begin{abstract}
We present observations in the H53$\alpha$ line and radio continuum at 43~GHz carried out 
with the VLA in the D array ($2''$ angular resolution) toward the starburst galaxy NGC~5253.
VLA archival data have been reprocessed to produce a uniform set of 2, 1.3 and 0.7~cm high 
angular ($0\rlap.{''}2 \times 0\rlap.{''}1$) radio continuum images. 
The RRL H53$\alpha$, a previously reported measurement of the H92$\alpha$ RRL flux density 
and the reprocessed high angular resolution radio continuum flux densities have been modeled using a collection of HII regions.
Based on the models, the ionized gas in the nuclear source has an electron density of $\sim 6 \times 10^4$~cm$^{-3}$ and 
an volume filling factor of 0.05.
A Lyman continuum photon production rate of $2 \times 10^{52}$~s$^{-1}$ 
is necessary to sustain the ionization in the nuclear region.
The number of required O7 stars in the central 1.5~pc of the supernebula is $\sim 2000$.
The H53$\alpha$ velocity gradient (10~km~s$^{-1}$~arcsec$^{-1}$) 
implies a dynamical mass of $\sim 3 \times 10^5$~M$_{\odot}$; this mass 
suggests the supernebula is confined by gravity. 

\end{abstract}


\section{INTRODUCTION}

\begin{figure}[!ht]
\plotone{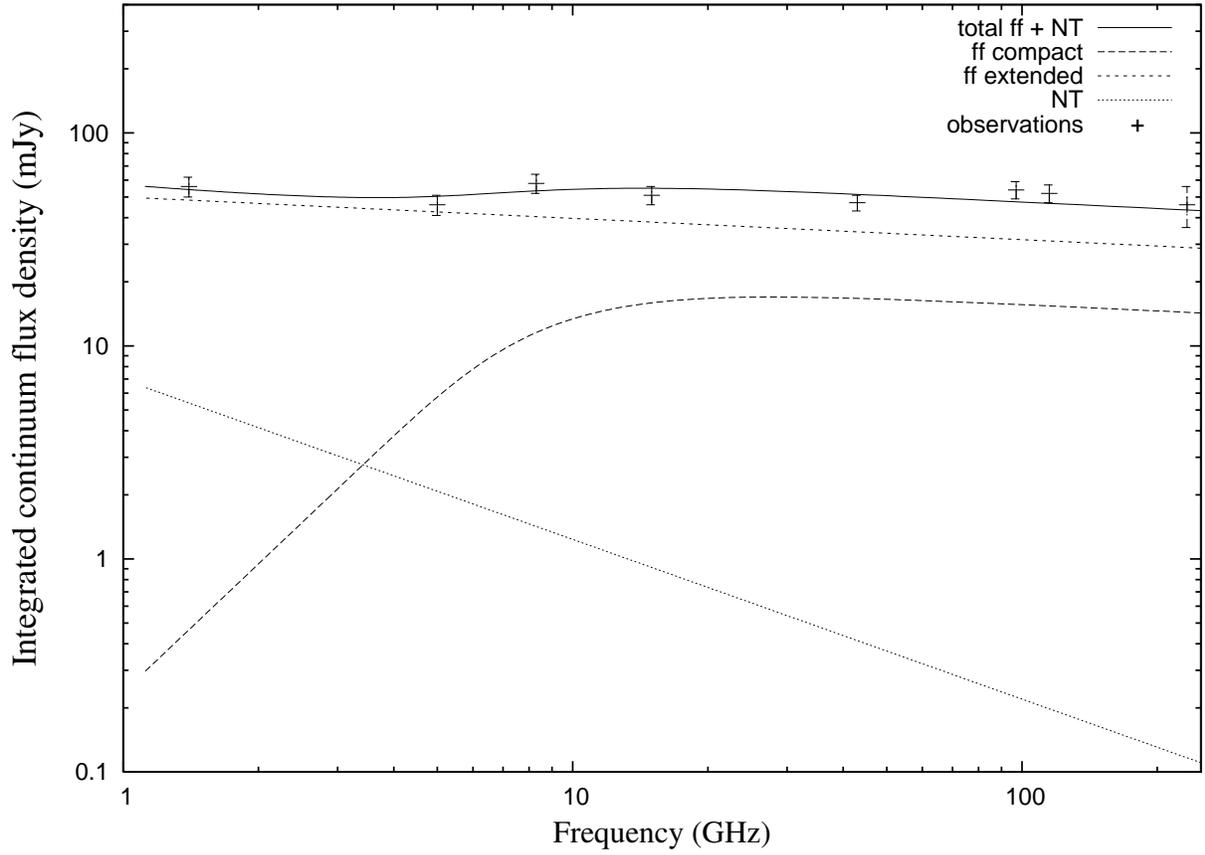}
\caption{Total radio continuum spectrum of NGC 5253 over the frequency range of 1 to 230 GHz.
The total values are determined over a region of $20'' \times 40''$ with low angular resolution $\ge 2''$. 
The long dashed line represents the optically-thick compact ($< 1.5$~pc, $0\rlap.{''}1$) component determined from high-angular resolution observations over the inner 
$0\rlap.{''}4$ (6~pc). 
The short dashed line is the extended ($> 1.5$~pc, $0\rlap.{''}1$) optically-thin component (S $\propto \nu^{-0.1}$).
The dotted line is the total synchrotron emission (S $\propto \nu^{-0.75}$) obtained from measurements at 5 GHz (Turner, Ho \& Beck~1998).
The solid line is the combined fit to the total observed continuum flux densities (listed in Table~2).}
\label{spectrum}
\end{figure}

NGC 5253 is a blue dwarf irregular galaxy (located at $\sim 4$~Mpc; Saha
et al.~1995) with an infrared luminosity of $\sim 2 \times 10^9$
L$_{\odot}$ (Beck et al. 1996). NGC~5253 hosts 
several groups of young star clusters, called super star clusters (SSC) (Gorjian 1996; Calzetti et al. 1997). 
The optical spectrum of NGC~5253 shows signatures of
large numbers of Wolf-Rayet (WR) stars (Campbell, Terlevich \&
Melnick et al.~1986; Walsh \& Roy~1989; Schaerer et al.~1997).
A nearly flat radio continuum spectra (see Fig.~\ref{spectrum})  also
indicates that free-free emission is the dominant emission mechanism and that NGC~5253 is a young starburst galaxy. 

The VLA observations made with similar angular resolution ($1''-2''$) at 6, 3.6 and 2~cm
(Beck et al.~1996 and Turner, Ho \& Beck~1998) reveal a complex structure in the central $20'' \times 40''$ region,
dominated by a compact source ($\leq 1''$; $1'' \sim 15$~pc). 
VLA observations of NGC 5253, made with higher angular resolution at 1.3 ($0\rlap.{''}33 \times 0\rlap.{''}12$) 
and 2~cm ($0\rlap.{''}22 \times 0\rlap.{''}08$) reveal 
two compact sources. The main source has a deconvolved angular size of
$0\rlap.{''}10 \times  0\rlap.{''}05$ which corresponds to $\sim 1-2$~pc (Turner, Beck \& Ho~2000).
NICMOS observations show the presence of a double cluster in the nucleus of the galaxy 
separated by $6-8$~pc ($\sim 0\rlap.{''}5$) \citep{Al04} which may be related with the double radio 
nebula detected by Turner, Beck \& Ho~(2000).
Higher angular resolution observations at 0.7~cm ($74 \times 17$~mas) show that the stronger component of the double nebula is compact,
but partially resolved at scales of $0\rlap.{''}05$ (Turner \& Beck 2004).
The main compact radio source (Turner, Ho \& Beck~1998) was named ``the supernebula''. 
Beirao et al.~(2006), based on mid-IR SPITZER observations from NGC 5253, report that mid-IR emission is dominated by 
an unresolved cluster, coincident with the compact supernebula. Beirao et al.~(2006) also observed an anticorrelation
between PAH strength and UV radiation, suggesting destruction of PAH molecules in the central region.

Based on 1.3 and 2~cm wavelength observations, Turner, Ho \& Beck~(2000) propose that the dominant radio
source is a strong candidate for a globular cluster in the process
of formation. The supernebula is visible at radio and infrared
frequencies with no obvious optical counterpart. This supernebula
is partially optically thick at 2~cm with an electron density of
$\sim 4 \times 10^4$~cm$^{-3}$ (Meier et al.~2002), typical of compact young ($\sim
10^5$~yr) HII regions in our galaxy (Wood \& Churchwell 1989).
Radio and infrared observations require a Lyman continuum flux of
$4 \times 10^{52}$~s$^{-1}$, equivalent to a few thousand O7
stars (Gorjian et al.~2001). The radio recombination line
H92$\alpha$ has been observed toward the supernebula and has been
interpreted as arising from gas around a SSC (Mohan et al.~2001).
Infrared Bracket $\alpha$ and $\gamma$ recombination line
observations revealed that the gas in the nebula is potentially bound by gravity
(Turner et al.~2003). Based on 0.7~cm wavelength observations, Turner \& Beck~(2004) confirm that the supernebula is a giant compact HII
region that is gravitationally bound. 
In order to reproduce the radio continuum emission, the nebula requires the excitation of $\sim 4000$ O7 stars 
within the central 5 pc region (Turner, Beck \& Ho 2000).

In this paper we present observations of the RRL H53$\alpha$
toward NGC 5253 using the VLA. The H53$\alpha$ RRL provides information 
of the physical properties of the optically thin ionized gas in the supernebula.
 In Section 2 the technical details
of the observations are presented. In Section 3 we present the
results, while in Section 4 we provide the discussion. In Section 5
we present the summary.

\section{VLA Observations.}

\subsection{H53$\alpha$ line.}

The H53$\alpha$ line ($\nu_{rest}=$42951.9~MHz) was observed in
the D configuration of the VLA on 2004, May 31 and June 02, 11, 12, and 13.
We used observation cycles with integration times of 10~min on NGC~5253 and 1~min
on the phase calibrator J1316-336 ($\sim 1.5$~Jy displaced by $6^{\circ}$). Two frequency
windows (LOs) were used to observe the RRL H53$\alpha$, centered at
42885.1 and 42914.9~GHz. For each frequency window,
the on-source integration time was $\sim 2$~hrs, using the mode of 15
spectral channels with a channel separation of 3.125~MHz ($\sim  22$~km~s$^{-1}$).
The data calibration was carried out for each frequency
window using the continuum channel, consisting of the central 75\%  of the band.
The flux density scales were determined from observations of J1331+305
(3C286; 1.47~Jy).
The bandpass response of the instrument was corrected using observations of J1337-129 ($\sim  9.5$~Jy).
The parameters of the observations are summarized in Table~1.
The coordinates listed in this Table~1 corresponds to the center of the observations.
In order to track reliably the phase variations introduced by
the troposphere, the calibration of the data was performed correcting for
the phases in a first step and subsequently correcting for both amplitude and phase.
The line data cubes were Hanning-smoothed to reduce the Gibbs effect and the final
velocity resolution is $\sim 44$~km~s$^{-1}$.
The line data were further calibrated using the solutions obtained by
self-calibrating the continuum channel of each frequency window.
The radio continuum images were obtained by combining the continuum
channels of each frequency window using the task DBCON from AIPS, and
the self-calibration method was also applied to this combined data.
The continuum emission was subtracted for each frequency window using the AIPS task
UVLSF with a zero order polynomial fit based on the line free channels.
The H53$\alpha$ line cubes and the 43 GHz continuum image were made
using a natural weighting scheme and then convolved to obtain a circular
Gaussian beam of $2\rlap.{''}0$
(P.A.$=0^{\circ}$). The combination of the different frequency
windows was carried out following a similar method to that used for the
H53$\alpha$ line observed toward M82 \citep{RoR04};
the two line cubes were combined into a single line cube after
regridding in frequency the line data for each frequency window.
This process was carried out using the GIPSY reduction package and a total of 19 velocity channels were obtained after putting together the two LO windows.
The total line bandwidth, after combining all the windows, is about
60~MHz (400~km~s$^{-1}$).

\subsection{Radio continuum at 2~cm, 1.3~cm and 0.7~cm}
In addition to the continuum emission at 0.7 cm, obtained from the line-free channels of the 
H53$\alpha$ line observations (at angular resolution of $2\rlap.{''}0$), we have included archive continuum data in the analysis.
VLA continuum observations at 2, 1.3 and 0.7 cm made with higher angular resolution ($< 1''$) toward NGC~5253 have been reprocessed 
from the VLA archive and used here to determine the properties of the continuum of the nuclear regions of NGC~5253 over angular scales 
 less than $7''$.
The observations at 2 and 1.3~cm were made with the A configuration of the VLA on 1998 April 9 (Turner, Beck \& Ho 2000).
The observations at 0.7~cm were made with the VLA in the A configuration including Pie Town on 2002 March 9 (Turner \& Beck~2004). 
The phase calibrator for the 0.7~cm is J1316-336 (J2000).
Absolute flux density calibration for all the 2~cm, 1.3~cm and 0.7~cm observations was based on observations of 3C286. 
The data at the three frequencies was brought to the same angular resolution by suitable choices of both
weighting in the uv plane and uv range selection. The final angular resolution of all three images is 
$0\rlap.{''}2 \times 0\rlap.{''}1$ (P.A.$=0^\circ$).

\section{RESULTS}

Figure~1 shows spectrum made with the low angular resolution ($\ge$ 2") observations summarized in Table 2. These low angular resolution
observations refer to the whole of the nucleus of NGC 5253. This spectrum is nearly flat in the frequency range 1-230 GHz. 
As noted before by Meier et al.~(2002), a single component that consists of optically thin free-free extended emission (S $\propto \nu^{-0.1}$) does not 
explain the observed flux densities at frequencies $\gtrsim 10$~GHz. 
An optically thick free-free component with an electron density
of $\sim 6 \times 10^{4}$~cm$^{-3}$ is necessary to explain the observed flux densities in the frequency range of $1-230$~GHz;
this optically thick free-free component has a turnover frequency at $\sim 9$~GHz.
The electron density value of the compact ionized gas component was obtained based on high angular resolution radio observations and
the fit is described in Section 4.2. 
There are at least three non-thermal components with angular sizes between $1''$ and $4''$,
which only contribute about 2~mJy at 6~cm and 1 mJy at 2~cm (see Turner, Ho \& Beck~1998) . 
Turner, Ho \& Beck~(1998) report three non-thermal sources based on observations at 2 and 6~cm, 
located near (east, southeast and northwest) the central super star cluster.
Thus, a third non-thermal (S~$\propto \nu^{-0.75}$) component with S$_{6cm}=2$~mJy  has also been used in 
the fit of the total continuum emission shown in Figure~1. 
The flux density level of the extended optically-thin component is fit to the flux density values obtained
from subtraction of the optically thick free-free and the synchrotron emission from the observed total flux densities.

\begin{figure}[!ht]
\epsscale{0.5}
\plotone{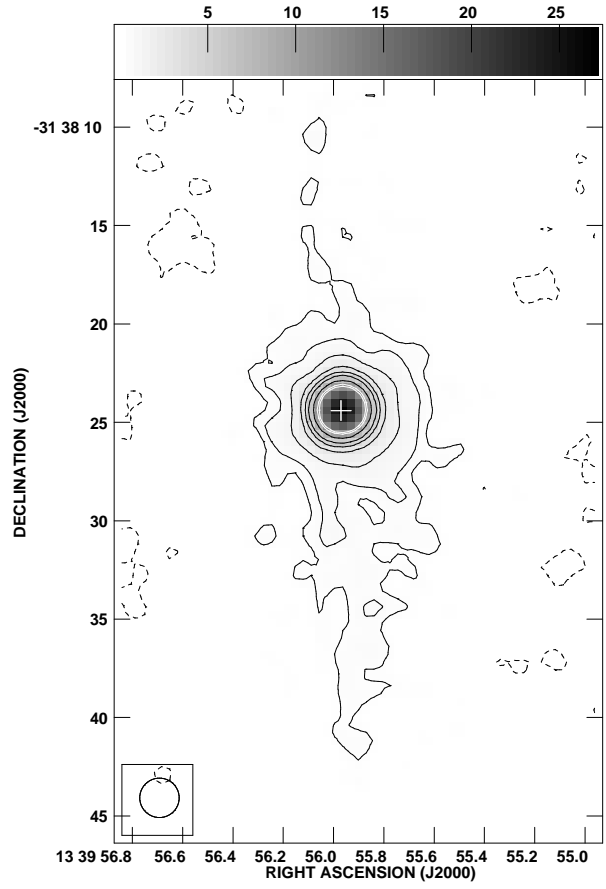}
\caption{Radio continuum image of NGC~5253 at 43~GHz obtained using the VLA in the D array. 
Contour levels are drawn at $-3$, 3, 5, 10, 20, 30, 40, 60, 80, and 90 times the rms of 0.13~mJy~beam$^{-1}$.
The cross shows the position of the 43~GHz continuum peak. The angular resolution is $2''$.}
\label{cont43}
\end{figure}

Figure~\ref{cont43} shows the 43 GHz radio continuum observations carried out in the D-array of the VLA (at $2''$ angular resolution).
The integrated 43~GHz continuum flux density measured at this angular resolution is $45 \pm 4$~mJy.
The radio continuum emission at 43 GHz extends $\sim 30''$ in the N-S direction and $\sim 10''$ in the E-W direction.
The extended emission along the N-S direction agrees with previous observations at 6, 3.6 and 2~cm \citep{Tu98}.
The central region dominates the continuum emission and has a deconvolved angular size of $\sim 1\rlap.{''}0$ at 43~GHz.
The position of the 43~GHz continuum peak ($\sim 27$~mJy~beam$^{-1}$) is $\alpha (2000)=13^{h} 39^{m} 55\rlap.{^s}97 \pm 0.1$,
$\delta (2000)=-31^{\circ} 38' 24\rlap.{''}4 \pm 0.1$, in agreement with previous observations at 2 and 1.3~cm \citep{Tu00}.
The position of the 43~GHz continuum emission peak is coincident with the position of the source labeled ``source A'' by \citet{Tu98}.

\begin{figure}[!ht]
\epsscale{0.7}
\plotone{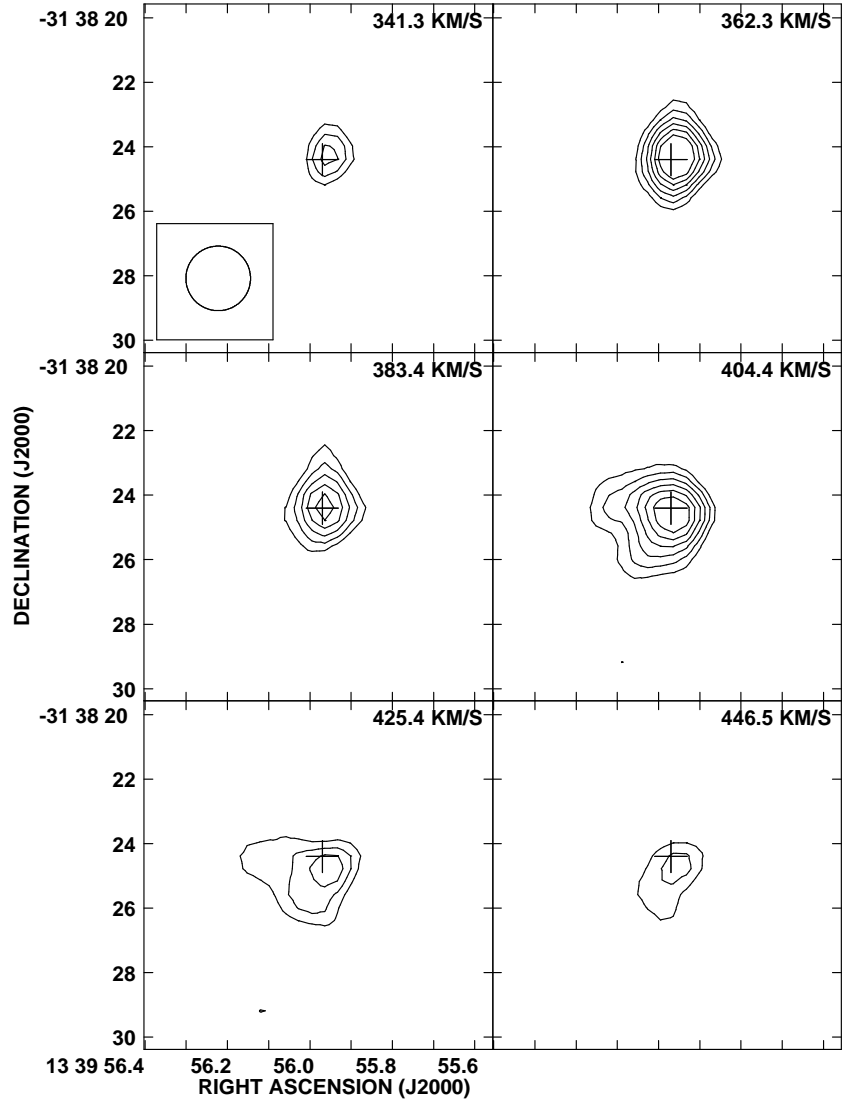}
\caption{ Channel images of the H53$\alpha$ line emission (contours) toward
NGC~5253 obtained using the VLA in the D array.
Contours are $-3$, 3, 4, 5, 6, 7, and 8 times  0.4~mJy~beam$^{-1}$, the rms noise.
The cross shows the position of the 43~GHz radio continuum peak. The synthesized beam ($2''$) is shown in the first panel.
The central heliocentric velocity is listed for each image.}
\label{chmaph53}
\end{figure}

\begin{figure}[!ht]
\plotone{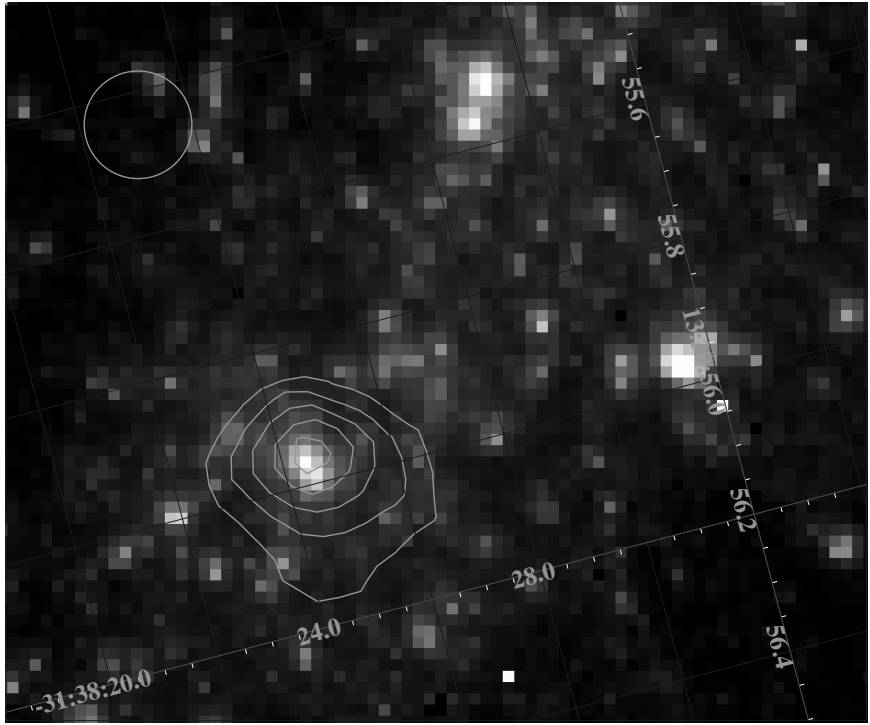}
\caption{ Overlay of the velocity integrated (moment 0) of the H53$\alpha$ line
emission (contours) toward NGC 5253 using the VLA in the D array and the H band image from NICMOS (F160W, $1.6~\mu m$) toward NGC 5253.
The NICMOS image has been shifted $\sim 1''$ to the NE (consistent with the astrometrical precision of HST) in order that the peak
of the H53$\alpha$ line emission coincides with the peak of the IR.
Contour levels are drawn at 5, 10, 20, 30, ..., 90, 95\% the peak line emission of 0.36~Jy~beam$^{-1}$~km~s$^{-1}$.
The cross shows the peak of the 43 GHz continuum emission.
The HPFW for the H53$\alpha$ line image is $2''$.}
\label{mom0}
\end{figure}

\begin{figure}
  \plotone{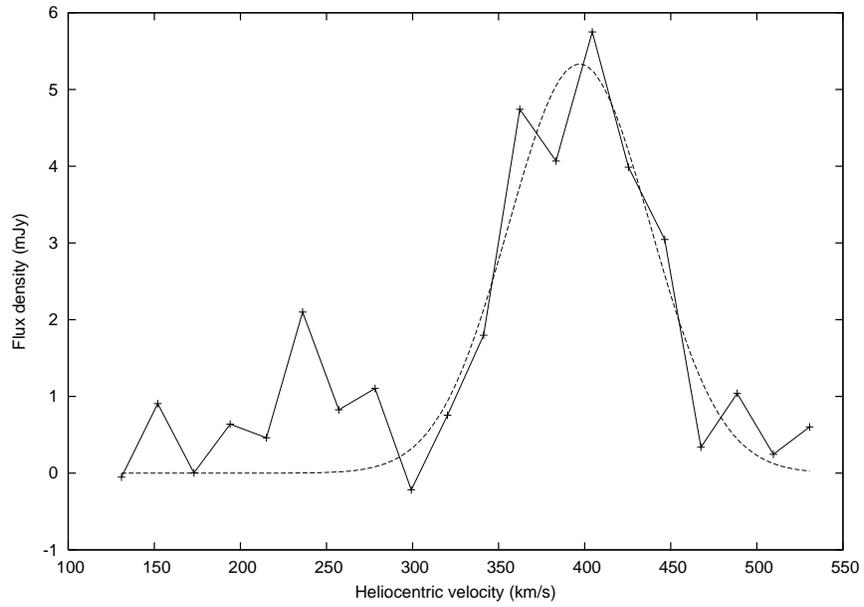}
\caption{Total integrated H53$\alpha$ spectrum from NGC~5253, obtained by integrating over the central diameter of $2''$.
The thick line shows the data and the dashed line shows the Gaussian fit. The rms in the velocity range $150-300$~km~s$^{-1}$  
is 0.45~mJy~beam$^{-1}$ and 0.25~mJy~beam$^{-1}$ in the velocity range $350-450$~km~s$^{-1}$.}
\label{profh53}
\end{figure}

Figure~\ref{chmaph53} shows the H53$\alpha$ line velocity-channel images of NGC 5253 at an angular resolution of $2''$.
The H53$\alpha$ line emission is observed in the heliocentric velocity range $\sim 340$ to 450~km~s$^{-1}$.
The peak H53$\alpha$ line flux density is $\sim 3.6$~mJy~beam$^{-1}$.
The deconvolved angular size measured in the H53$\alpha$ line is $\sim 1\rlap.{''}0$.
We have obtained the $1.6\mu$m NICMOS image from the HST archive in order to compare the H53$\alpha$ RRL with the IR emission; 
the plate scale of the NICMOS observations is $0\rlap.{''}034$.
Following Turner et al.~(2003) we have assumed that the brightest source in the $1.6\mu$m NICMOS can be identified with 
the radio peak as observed at an angular resolution of $2''$. We thus have shifted the NICMOS image by $\sim 1''$ in the NE direction 
in order that the IR and radio peaks coincide, consistent with the astrometrical precision of HST. 
Figure~\ref{mom0} shows the overlay of the integrated H53$\alpha$ line emission on the $1.6\mu$m 
after shifting the IR image.
The integrated H53$\alpha$ line ($0.31 \pm 0.03$~Jy~km~s$^{-1}$), 
as obtained from a gaussian fit, has a peak line flux density
$S_L=5.3 \pm 0.6$~mJy, a FWHM of $58 \pm 12$~km~s$^{-1}$ and a central heliocentric velocity of 
$397 \pm 5$~km~s$^{-1}$.
The velocity integrated H53$\alpha$ line emission is shown in Figure~\ref{profh53}.

\section{DISCUSSION}

\subsection{Total radio continuum emission}

The spatially integrated radio continuum flux density obtained with low angular resolution, $S_{43GHz} \simeq 45$~mJy is
comparable with previous measurements from interferometric observations carried out in the wavelength range of 21 to 0.13~cm ($1.4-230$~GHz).
Previous observations have revealed that the radio continuum emission at 43 GHz is dominated by thermal free-free emission (Beck et al.~1996)  
and our 43~GHz radio continuum observations suggest that the contribution from non-thermal emission is negligible at wavelengths shorter than 6~cm ($> 5$~GHz, see Figure~1). 
Thus, no contribution from synchrotron sources will be considered in the radio continuum emission models (see Section 4.2).

\subsection{Models of a collection of HII region}
Models that consist of HII regions have been used to estimate the electron density of the ionized gas in
starburst galaxies like M82 (Rodr\'{\i}guez-Rico et al.~2004), NGC 253 (Rodr\'{\i}guez-Rico et al.~2006) and Arp 220 
(Anantharamaiah et al. 2000; Rodr\'{\i}guez-Rico et al.~2005).
The models are constrained by: 
1) the physical diameter of the region associated with the RRLs emission and 
2) the measured radio continuum and RRL flux densities.
The RRLs H92$\alpha$ (Mohan et al. 2001) and H53$\alpha$ are used, along with the high-angular resolution radio continuum 
observations described in Section~2.2, to estimate the electron density of the ionized gas in NGC~5253.

Based on the high angular resolution VLA archival observations ($0\rlap.{''}2 \times 0\rlap.{''}1$), described in Section 2.2 
the deconvolved angular size of the compact nuclear region is $\sim 0\rlap.{''}1$ ($\sim 1.5$~pc).
The peak position measured in these high angular resolution radio continuum images agrees with the peak position of
the RRLs H53$\alpha$ and H92$\alpha$ (Mohan et al. 2002). Thus, it is assumed that the H53$\alpha$ and the H92$\alpha$ RRLs 
arise from the same $\sim1.5$~pc ($\sim 0\rlap.{''}1$) region.
The radio continuum flux densities, at $0\rlap.{''}2 \times 0\rlap.{''}1$ angular resolution,
are S$_{2cm}=9.3 \pm 1$~mJy, S$_{1.3cm}=11 \pm 1$~mJy and S$_{0.7cm}=9.5 \pm 1$~mJy. 
The velocity integrated H53$\alpha$ and H92$\alpha$ line emission are 50~mJy~km~s$^{-1}$ and 203~mJy~km~s$^{-1}$, respectively.
Acceptable models are those that can reproduce the continuum flux densities
and the velocity integrated line emission on both RRLs. 
Based on the size measured in the radio continuum images, the maximum volume 
that this collection of HII regions may occupy is the volume of a spherical 1.5~pc region.

Following the procedure listed in Rohlfs et al.~1996, we use models that consist of HII regions ionized by O7 early-type 
stars ($10^{49}$~Lyman continuum photons~s$^{-1}$).  
The emission measure of each HII region is $EM_C=2 n_e^2~l$.
The continuum optical depth is (Altenhoff et al.~1960)
\begin{equation}
\tau_c=8.235 \times 10^{-2}~\bigg(\frac{T_e}{K}\bigg)^{-1.35}~\bigg(\frac{\nu}{GHz}\bigg)^{-2.1}~\bigg(\frac{EM}{cm^{-6}~pc}\bigg),
\end{equation}
\noindent
and the black body radiation is defined by 
\begin{equation}
\frac{B_\nu}{mJy} = 3.07 \times 10^{7}~\bigg(\frac{\nu}{GHz}\bigg)^2~\bigg(\frac{T_e}{K}\bigg).
\end{equation}

The electron temperature of the thermally ionized gas is assumed to be T$_e = 10^4$~K and $\nu$ is the frequency.
The radio continuum flux density (S$_{cth}$) for each HII region in the models is given by:
\begin{equation}
\frac{S_{cth}}{mJy}=\bigg(\frac{B_\nu}{mJy}\bigg)~\bigg(\frac{\Omega}{sterad}\bigg)~(1-e^{-\tau_c}),
\end{equation}
where $\Omega=\pi(l/4.0~Mpc)^2$ is the solid angle subtended by the HII region.

The models for RRL emission take into account deviations from local thermodynamic equilibrium (LTE) using 
the departure coefficients $b_n$ and $\beta_n$. The peak line flux density is:
\begin{equation}
\frac{S_L}{mJy}=\bigg(\frac{B_\nu}{mJy}\bigg)~\bigg(\frac{\Omega}{sterad}\bigg)~\Big[ \bigg( \frac{\tau_c + b_n \tau_L^*}{\tau_c + \tau_L}\bigg)~(1-e^{-(\tau_c+\tau_L)})
-(1-e^{-\tau_c}) \Big],
\end{equation}
where $\tau_L^*= (\tau_c~r)$ is the line optical depth in LTE and $\tau_L= (\tau_c~r~b_n~\beta_n$) is the non LTE line optical depth.
The line-to-continuum ratio $r$ defined by: 
\begin{equation}
r=2.33 \times 10^4 \bigg( \frac{\Delta \nu}{kHz}\bigg)^{-1.0} \bigg(\frac{\nu}{GHz} \bigg)^{2.1} \bigg( \frac{T_e}{K} \bigg)^{-1.15} \bigg( \frac{EM_L}{EM_C} \bigg).
\end{equation}
In this equation the line width is $\Delta \nu$ and $EM_L=0.9 EM_C$ is the emission measure of the line, 
assuming a fractional number abundance of He$^{+}$ to H of 0.1.

\begin{figure}[!ht]
\plotone{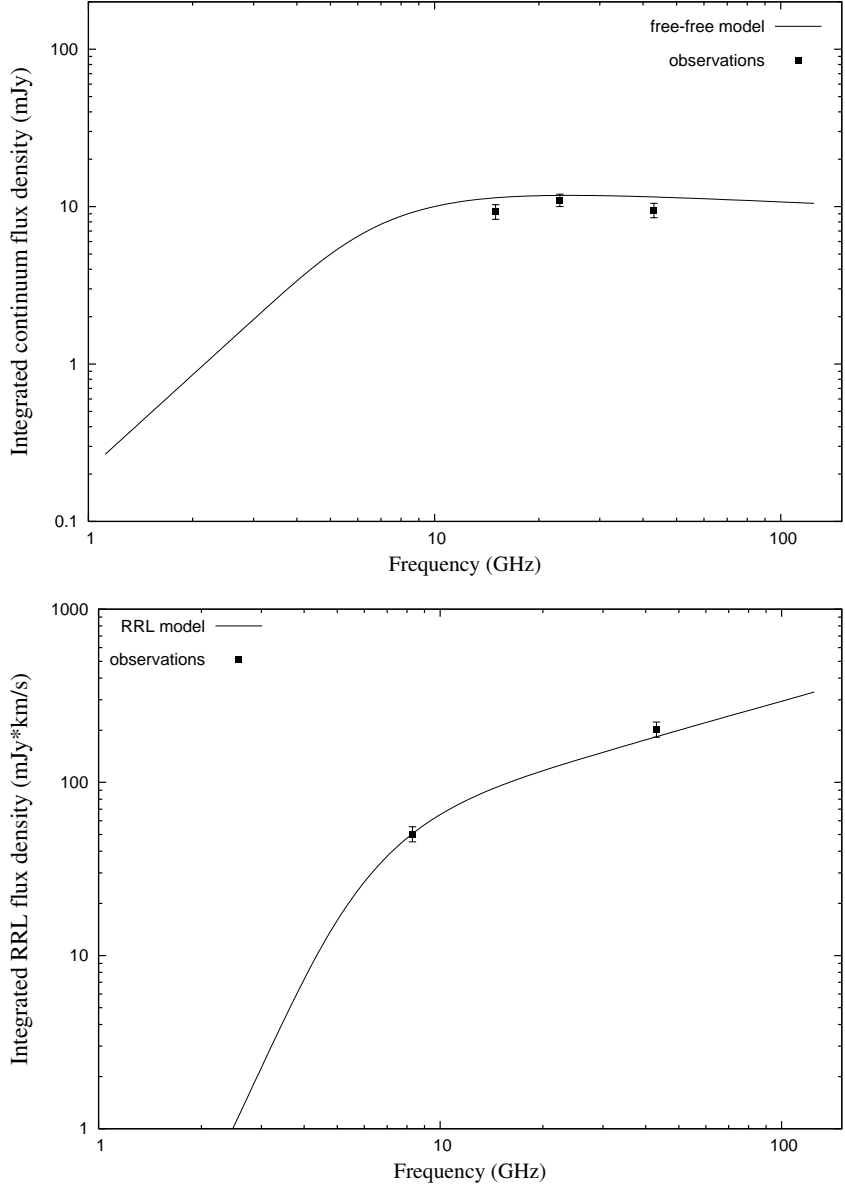}
\caption{Single density (n$_e=6 \times 10^4$~cm$^{-3}$) models for the continuum (top) and RRL (bottom) emission 
from the supernebula are shown as solid lines. 
Data points in the continuum and RRL are marked by the filled squares. 
The models consist of a collection of HII regions with electron temperature of $10^4$~K. 
The radio continuum observations consist of the reprocessed archival data with $0\rlap.{''}2 \times 0\rlap.{''}1$ angular resolution (listed in Table~2).
The RRL observations are from this paper (H53$\alpha$) and from Mohan et al.~(2001) (H92$\alpha$).
These recombination lines arise from the compact ($\sim 1.5$~pc) radio supernebula.} 
\label{models}
\end{figure}

The electron density of the ionized gas in these models was varied between $10^2 - 10^6$~cm$^{-3}$.
Figure~\ref{models} shows the results from single-density models that fit the radio continuum and RRL observations for NGC~5253.
The solution for the electron density obtained from the models is $\sim 6 \times 10^4$~cm$^{-3}$.
This electron density is comparable to the derived value of $4-5 \times 10^4$~cm$^{-3}$ (Turner, Beck \& Ho~2000 and Meier et al.~2002).
The total volume filling factor of the dense ionized gas is $\sim 0.05$ and the total area filling factor is $\sim 0.95$.
Based on the results of the single density models, the Lyman continuum 
needed to ionize the~1.5 pc region is $\sim 2 \times 10^{52}$~s$^{-1}$, corresponding to $\sim 2000$ O7 stars.
The total mass of O7 stars, assuming 20~M$_{\odot}$ for each O7 star (Herrero et al. 1992), is about $4 \times 10^4$~M$_{\odot}$.

Comparison of the velocity integrated H53$\alpha$ line emission and NICMOS images at 1.6~$\mu$m (Figure 4) 
suggests that the radio supernebula has a close counterpart in infrared. Previous observations 
revealed the radio supernebula is associated with the brightest Brackett line source in NGC 5253 (Turner et al.~2003).
The dust extinction in the infrared at 1.6~$\mu$m is evident (Figure 4) as
the NICMOS image reveals a very compact emission toward the supernebula compared to the radio image in H53$\alpha$ integrated line emission.
Based on the velocity field, a velocity gradient of $\sim 10$~km~s$^{-1}$~arcsec$^{-1}$ is observed,
suggesting an apparent rotation of the ionized gas in the supernebula; the nearly parallel isovelocity contours 
suggest rotation in the supernebula (see Figure~7). 
The dynamical mass  M~$sin~i$ in the supernebula, with physical size of $\sim 1.5$~pc, is $\sim 3 \times 10^5$~M$_{\odot}$. 
This mass is comparable to the mass estimated for the nebula to be in virial equilibrium ($\sim 4 \times 10^5$~M$_{\odot}$, Turner \& Beck 2004). 
\citet{Tu04}, based on the H92$\alpha$ and the Br$\gamma$ line FWHM ($\sim 75$~km~s$^{-1}$), 
concluded that the supernebula is gravity bounded. 
The mass implied by the velocity gradient is consistent with the luminosity implied by starburst models with this N$_{Lyc}$ (Leitherer et al.~1999).
The H53$\alpha$ line FWHM of $58 \pm 5$~km~s$^{-1}$ confirms that the supernebula is gravity bounded.
The difference in the line widths is interpreted as contribution of the extended ionized gas component to the Br$\gamma$ and H92$\alpha$ lines compared to 
the H53$\alpha$ line that traces mostly the compact ionized gas component.

\section{CONCLUSIONS}

The H53$\alpha$ line and the radio continuum emission at 43~GHz were observed with the VLA in the D array toward the galaxy NGC~5253.
VLA archival data have been reprocessed to produce a uniform set of 2, 1.3 and 0.7~cm 
continuum images with angular resolution of $0\rlap.{''}2 \times 0\rlap.{''}1$.
Using a single density model to reproduce the observed radio continuum at 2, 1.3 and 0.7 cm as well as the H53$\alpha$ and H92$\alpha$ 
RRL emission in the central compact region ($\sim 1.5$~pc), the average electron density of the ionized gas is $\sim 6 \times 10^4$~cm$^{-3}$.
The 43~GHz radio continuum emission arise from an optically thin free-free emission source. 
The Lyman continuum photons rate of $\sim 2 \times 10^{52}$~s$^{-1}$ is necessary to sustain the ionization in the 1.5~pc
in the inner region of the galaxy NGC~5253, corresponding to $\sim 2000$~O7 stars.
The peak of the H53$\alpha$ line emission coincides with the brightest source observed in the 1.6$\mu$m IR image within $1''$.
The velocity gradient measured in the H53$\alpha$ line (10~km~s$^{-1}$~arcsec$^{-1}$), if interpreted as rotation, 
implies a dynamical mass of $\sim 3 \times 10^5$~M$_{\odot}$ within the central 20~pc.
The rough agreement of the mass derived from the apparent rotation with the mass estimated from the  
observed Lyman continuum rate, N$_{Lyc}$, suggests that the
gas motion within the supernebula is governed by gravity.

The National Radio Astronomy Observatory is a facility of the National
Science Foundation operated under cooperative agreement by Associated
Universities, Inc. CR and YG acknowledge support from UNAM and
CONACyT, M\'exico.

\clearpage

\begin{deluxetable}{cc}
\tablecolumns{2}
\tablewidth{0pc}
\tablecaption{Observing parameters for NGC~5253 using the VLA. }
\tablehead{ \colhead{Parameter} & \colhead{H53$\alpha$ RRL (43 GHz)}}
\startdata
Right ascension (J2000) \dotfill & 13$^h$39$^m$$55\rlap.{^s}$96\\
Declination (J2000)\dotfill & $-31 ^{\circ} 38' 24\rlap.{''}38$\\
Angular resolution continuum \& line \dotfill & $2'' \times 2''$, P.A.$=0^{\circ}$ \\
On-source observing duration (hr)\dotfill & 8  \\
Bandwidth (MHz)\dotfill & 60\\
Total number of spectral channels\dotfill & 19\\
Optical Heliocentric Velocity V$_{Hel}$ (km~s$^{-1}$)\dotfill & 397   \\
Velocity coverage (km~s$^{-1}$)\dotfill & 400 \\
Velocity resolution (km~s$^{-1}$)\dotfill & 44 \\
Amplitude calibrator\dotfill & J1331+305\\
Phase calibrator\dotfill & J1316$-$336\\
Bandpass calibrator\dotfill & J1337$-$129\\
RMS line noise per channel (mJy/beam)\dotfill & 0.4\\
RMS, continuum (mJy/beam)\dotfill & 0.2\\
\enddata
\end{deluxetable}

\clearpage

\begin{deluxetable}{cccc}
\tablecolumns{2}
\tablewidth{0pc}
\tablecaption{Radio continuum flux densities for NGC~5253. }
\tablehead{ \colhead{Wavelength} & \colhead{Flux density} & Angular resolution & Reference\\
 \colhead{(cm)} & \colhead{(mJy)}} 
\startdata
20   \dotfill  & $56  \pm 1$  & $9'' \times 4''$ \tablenotemark{a} & 1 \\              
6    \dotfill  & $46  \pm 1$  & $2\rlap.{''}2 \times 1\rlap.{''}4$  \tablenotemark{a}  & 1 \\
3.6  \dotfill  & $40  \pm 6$  & $2\rlap.{''}2 \times 1\rlap.{''}4$  \tablenotemark{a}  & 2 \\
2    \dotfill  & $51  \pm 1$  & $2\rlap.{''}2 \times 1\rlap.{''}4$  \tablenotemark{a}  & 1 \\
               & $9.3 \pm 1$  & $0\rlap.{''}2 \times 0\rlap.{''}1$  \tablenotemark{b}  & 3 \\
1.3  \dotfill  & $11  \pm 1$  & $0\rlap.{''}2 \times 0\rlap.{''}1$  \tablenotemark{b}  & 3 \\
0.7  \dotfill  & $47  \pm 4$  & $2''$                               \tablenotemark{a}  & 3 \\
               & $9.5 \pm 1$  & $0\rlap.{''}2 \times 0\rlap.{''}1$  \tablenotemark{b}  & 3 \\
0.31 \dotfill  & $54  \pm 5$  & $9\rlap.{''}0 \times 6\rlap.{''}8$  \tablenotemark{a}  & 5 \\
0.26 \dotfill  & $52  \pm 4$  & $14\rlap.{''}0 \times 6\rlap.{''}5$ \tablenotemark{a}  & 4 \\
0.13 \dotfill  & $46  \pm 10$ & $6\rlap.{''}5 \times 4\rlap.{''}5$  \tablenotemark{a}   & 5 \\
\enddata
\tablenotetext{a} {Values obtained by integrating over the central $20'' \times 40''$ 
region and used to obtain the total radio continuum spectrum (Figure~1). All values were obtained from the literature except the value at 0.7~cm.}
\tablenotetext{b} {Values were obtained from the reprocessed VLA data (see Section 2.2) integrating over the inner $0\rlap.{''}4$. These values
were used in the models of a collection of HII regions (Figure~6).}
\tablerefs{ (1) Turner, Ho \& Beck~(1998), (2) Mohan, Anantharamaiah \& Goss~(2001),(3) This paper, (4) Turner, Beck, \& Hurt~(1997), (5) Meier, Turner \& Beck~(2002).}
\end{deluxetable}

\clearpage

\clearpage

\clearpage

\clearpage

\clearpage

\clearpage

\clearpage

\begin{figure}[!ht]
\plotone{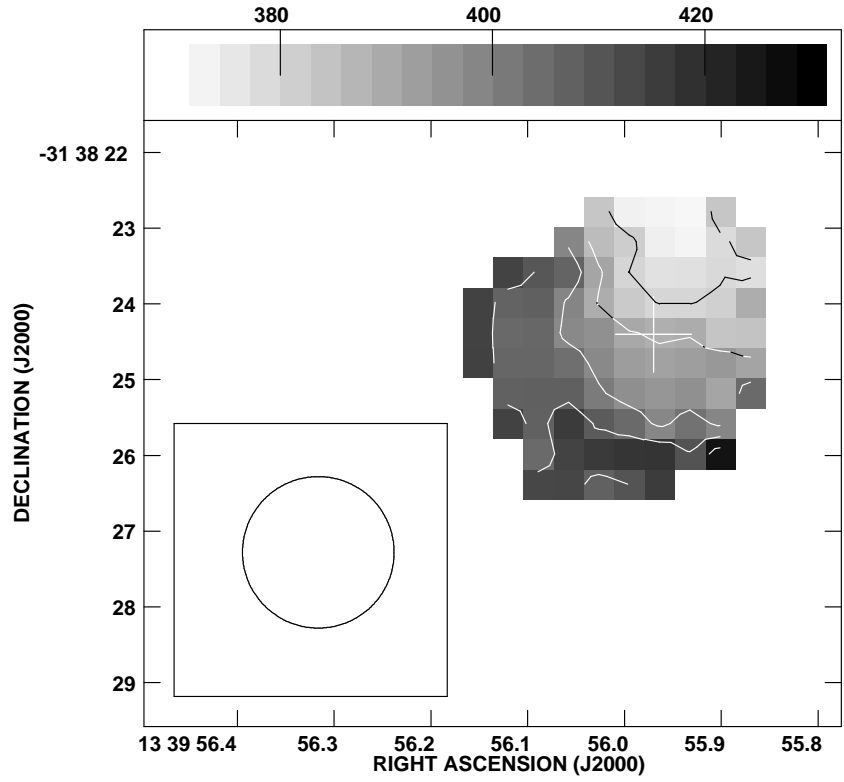}
\caption{ Velocity field observed in the H53$\alpha$ line from NGC~5253.
Contour levels are the heliocentric velocities at 380, 390, 400, 410 and 420~km~s${^{-1}}$.
The grayscale shows the heliocentric velocity field image in the H53$\alpha$ line.
The cross shows the peak of the 43 GHz continuum emission. The HPFW is  $2''$.}
\label{mom1}
\end{figure}

\end{document}